\documentclass[10pt,conference]{IEEEtran}

\usepackage[english]{babel}										%
\usepackage[protrusion=true,expansion=true]{microtype}				%
\usepackage{amsmath,amsfonts}					%
\usepackage{graphicx}									%
\usepackage{epstopdf}												%
\usepackage{booktabs}												%
\usepackage[square,sort,comma,numbers,compress]{natbib}

\usepackage{algorithm}
\usepackage{algorithmic}

\usepackage{listings}
\usepackage[usenames,dvipsnames]{xcolor}
\definecolor{listinggray}{gray}{0.5}
\definecolor{lbcolor}{rgb}{0.9,0.9,0.9}
\definecolor{darkgreen}{rgb}{0.0, 0.2, 0.13}
\usepackage[hang, small, labelfont=bf, textfont=it, up]{caption}	%
\usepackage[skip=0cm,list=true]{subcaption}
\usepackage{paralist}
\usepackage{flushend}
\usepackage{balance}

\let\oldenumerate\enumerate
\renewcommand{\enumerate}{
  \oldenumerate
  \setlength{\itemsep}{3pt}
  \setlength{\parskip}{0pt}
  \setlength{\parsep}{10pt}
}

\title{Implementation and Evaluation of Data Center Congestion Controller with Switch Assistance}					%

\author{Ahmed M. Abdelmoniem and Brahim Bensaou\\
Department of Computer Science and Engineering\\
The Hong Kong University of Science and Technology\\
Clear Water Bay, Hong Kong\\ 
\{amas, brahim\}@cse.ust.hk
}

\def\Rwnd{\textit{Rwnd }}

\makeatletter

\begin{document}

\newcommand{\SWITCH}[1]{\STATE \textbf{switch} (#1)}
\newcommand{\ENDSWITCH}{\STATE \textbf{end switch}}
\newcommand{\CASE}[1]{\STATE \textbf{case} #1\textbf{:} \begin{ALC@g}}
\newcommand{\ENDCASE}{\end{ALC@g}}
\newcommand{\CASELINE}[1]{\STATE \textbf{case} #1\textbf{:} }
\newcommand{\DEFAULT}{\STATE \textbf{default:} \begin{ALC@g}}
\newcommand{\ENDDEFAULT}{\end{ALC@g}}
\newcommand{\DEFAULTLINE}[1]{\STATE \textbf{default:} }

\maketitle

\begin{abstract}
In this work, we provide the design and implementation of a switch-assisted congestion control algorithm for data center networks (DCNs). In particular, we provide a prototype of  the switch-driven congestion control algorithm and deploy it in a real data center. The prototype is based on few simple modifications to the switch software. The modifications imposed by the algorithm on switch is to enable the switch to modify the TCP receive-window field in the packet headers. By doing so, the algorithm can enforce a pre-calculated (or target rate) to limit the sending rate at the sources. Therefore, the algorithm requires no modifications to the TCP source or receiver code which considered out of the DCN operators' control (E.g., in public cloud where the VM is maintained by the tenant). This paper describes in detail two implementations, one as a Linux kernel module and the second as an added feature to the well known software switch, Open vSwitch. Then we present evaluation results based on experiments of the deployment of both designs in a small testbed to demonstrate the effectiveness of the proposed technique in achieving high throughput, a good fairness and short flow completion times for delay-sensitive flows\footnote{Manuscript is accepted in  IEEE IPCCC 2015. This work is supported in part under Grants: HKPFS PF12-16707, REC14EG03 and FSGRF13EG14}. 
\end{abstract}

\begin{IEEEkeywords}
Congestion Control, Data Center Networks, Flow Control, Implementations, Linux NetFilter, Open vSwitch
\end{IEEEkeywords}

\section{Introduction}
\label{sec:1}
Data center (DC) environments are used to host a large number of different applications with heterogeneous traffic characteristics and performance requirements. These applications result in a mix of small but time-sensitive traffic flows such as control-messages or web-search traffic (called in the sequel mice) and long lived throughput-inclined flows such as VM-migration (called hereafter elephants).  Mice and elephant flows often co-exist within the same DC Network (DCN). Unlike the Internet, DCNs which are characterized by a large bandwidth and small round-trip delay, impose many new challenges to the overlying congestion control mechanisms of TCP, mainly because  TCP congestion control mechanism in all its variations  
\begin{inparaenum}[\itshape i)\upshape]
\item is unaware to the overlying application performance requirements and traffic characteristics; and 
\item is typically designed to achieve stability and high bandwidth utilization, while only targeting in practice fairness for long live elephant flows under ideal conditions\footnote{Mice flows often do not have enough traffic to achieve their traditional TCP fair share under TCP congestion control}.
\end{inparaenum}

Unlike the Internet that relies on routers with large buffers, DCNs use switches with small buffers (few megabytes to few hundred megabytes). Round trip delays in DCNs are short (few microseconds to few hundred microsecond) with a relatively large bandwidth (1Gbps to 10Gbps)  \cite{Phanishayee2008,Vasudevan2009,Chen2009,Alizadeh2010}. When such networks are fed with a mixture of mice and elephants, several congestion phenomena that cannot be simply inferred from packet losses and/or delay take place \cite{Chen2009,Alizadeh2010,Ahmed-GLOBECOM-2015}: 
\begin{inparaenum}[\itshape i) \upshape]
\item Incast traffic congestion: In partition/aggregate applications (e.g., MapReduce) many synchronized mice flows compete to exit from the same congested output port of a switch over a very short period of time. In the presence of small switch buffers they lead to excessive packet drops and timeouts; and,
\item Queue-buildup: the normal behaviour of TCP is to consume the bandwidth-delay product of the network including the buffer space in the routers and switches. Elephant flows in particular last long enough to increase their sending window to achieve this; as a result mice flows arriving to such bloated buffers experience repeated packet drops stretching their completion times unnecessarily due to timeouts.
\end{inparaenum}

Due to the impact and severity of these congestion symptoms on cloud users' experience, much recent research work has been devoted to addressing the shortcomings of TCP in DCNs. These works can in general be categorized into two categories: window based schemes (e.g., \cite{Alizadeh2010,Wu2013}) and fast loss recovery schemes (e.g., \cite{Vasudevan2009, Cheng2014}). 

The major drawback common to all these schemes is their reliance on partial changes to the sender/receiver TCP stack or for some a total replacement of TCP by a new protocol. In practice, while many tenants adopt ready-made VM images from the DCN provider, many elect to upload their own VM image, others elect to modify or fine tune the parameters of their VM network stack, and so on. In a nutshell one cannot assume all VM congestion control mehcanisms in the DCN are homogeneous, and as result, these solutions, despite their effectiveness, turn out to be limited to private DCNs where the TCP protocol is under full control of the DCN provider. To cope with this limitation we advocate a flow-aware approach similar to traditional flow-based system like ATM-ABR or XCP \cite{Katabi2002}, where congestion control in a DCN is treated as a network problem rather than a transport problem. In this perspective we proposed in \cite{Ahmed-CLOUDNET-2015} a mechanism called \textsl{Receiver window queue (RWNDQ)} that proved to be able to achieve a high efficiency for mice and elephants by keeping a low queue occupancy as well as a good fairness in both short and long term. We have also analysed the stability of RWNDQ mechanism using a simple analytical model and examined its effectiveness through ns2 simulations comparing it to TCP, XCP and DCTCP \cite{Ahmed-CLOUDNET-2015}. Many challenges remain to deploy such mechanism in real systems, the most important one being, how to implement such flow-awareness in the flow-averse IP environment, while maintaining TCP sender/receiver code untouched and without storing per flow states in the switches. In our approach, we track the number of active flows on each switch port by counting SYN-ACK/FIN TCP packets and rely on the TCP receiver-window field in the TCP header to convey the fair-share back to the sources. TCP flow control being a fundamental part of any TCP variant, our proposed mechanism fits-in without any change to the sender nor the receiver. We discuss in this paper how we can achieve this and implement this algorithm effectively in a real system with the following contributions:
\begin{itemize}
\item  We describe the implementation of RWNDQ as a run-time loadable Linux kernel module that can be used as a standalone buffer management mechanism in software switches or switches running Linux Network OS\footnote{For example, Pica8 pronto-3295: http://www.pica8.com/documents/pica8-datasheet-48x1gbe-p3290-p3295.pdf}. 
\item We also describe the implementation of RWNDQ as a new added feature to the data-path of the well-known industry standard software switch Open vSwitch (OvS)\footnote{``Open Virtual Switch Project'' http://openvswitch.org/} which is widely used in virtualized data centers. 
\item Finally, we build a small-scale testbed with 12 servers, Gigabit-ethernet switches and OvS on all the servers and in the core switch to study the performance of RWNDQ with multiple bottleneck links, incast traffic and buffer-bloating at both the ToR switches and core switches. 
\end{itemize}
Our results show that RWNDQ handles congestion gracefully, and is able to reconcile mice and elephants by enabling the former to finish their flows quickly and the latter to achieve a high throughput and fairness, even when the network is severely congested.

The remainder of the paper is organized as follow, we first briefly review RWNDQ in Section \ref{sec:2}. In Section~\ref{sec:3}, we discuss the implementation aspects of the Linux kernel module and show some experimental results. We then show how we added the RWNDQ as a feature of OvS and discuss testbed performance results in Section \ref{sec:4}. We discuss the related work in \ref{sec:5} and finally conclude the paper in \ref{sec:6}.

\section{RWNDQ Algorithm}
\label{sec:2}

\begin{figure}[ht]
	\centering
		\includegraphics[scale=0.95]{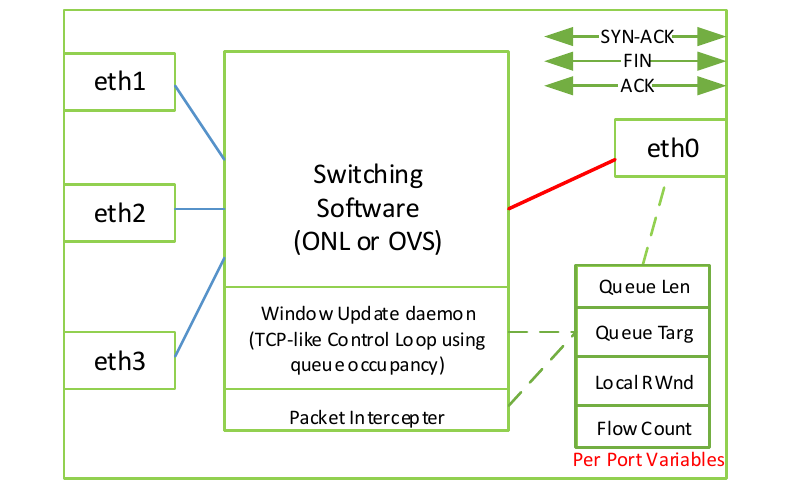}
	\caption{RWNDQ high level system-view design}
	\label{fig:rwndq}
\end{figure}

RWNDQ system overview is given in Figure~\ref{fig:rwndq}. RWNDQ is an algorithm that tries to maintain the queue at the switches below a given target by adjusting the senders rates. It runs in any switching point and maintains several state variables per switch port. It intercepts three types of packets and updates the state information per port accordingly: 
\begin{inparaenum}[\itshape i) \upshape]
\item SYN-ACK: which establishes a TCP flow on both directions (forward and backward) increments the flow count associated with the output queues of both directions by 1; 
\item FIN: which terminates a TCP flow on one direction (incoming direction only) decrements the associated flow count of that queue by 1; and
\item ACK: which is examined as a candidate to carry back in the receiver window field (\textit{Rwnd}) a window update based on a (switch) local window value of the reverse direction output queue.
\end{inparaenum}

RWNDQ is an event-driven scheme that deals with two major events: packet arrival, and window update timer expiry events as follows:  

\subsection{Window update daemon} 
As part of the initialization step or if this is the first flow (based on intercepted SYN-ACK), then the current local window is initially set to the target-queue occupancy worth of bytes then, because initially the aggregate bandwidth-delay product is unknown to the switch, RWNDQ enters a slow-start phase to start probing for the corresponding window size. Slow-start phase is terminated as soon as the current queue occupancy exceeds the predetermined queue target. As shown in listing~\ref{code:updater}, RWNDQ implements a local TCP-Like window control loop that tracks at regular intervals the deviation of the current queue occupancy from the target one. It calculates a ratio of current queue over target which directly controls the accumulated fraction of segments (MSS) added to or subtracted from the window update variable. The algorithm waits for a number of successive updates after which the current value of the per-port local window is updated with the value of update variable (except during slow start where it adds two MSS to the window). Note that RWNDQ waits for a number of accumulated updates before it is used to update the actual value of \Rwnd that is conveyed to the TCP sender. This enables a highly accurate estimation of the increment, while keeping the number of \Rwnd rewrites in the packet header reasonably small.

\small
\begin{lstlisting}[caption=RWNDQ Window Update Daemon, label=code:updater]
//Window Update Daemon
hrtimer_restart timer_callback(hrtimer *timer)
{
  timerrun=false;//assume no active ports
	//update the local window of all active ports
  net_device * dev=first_net_device(&init_net);
  i=0;
  while (dev!=NULL && i<devcount)
  {
     if(dev->ifindex==devind[i] && conncount[i]>0)
     {
		   timerrun=true;
			 backlog=dev->qdisc->qstats.backlog;
			 //target is set to 25\% of buffer length
			 //left shift by 2 => divide by 4 => 25\%
			 target=(dev->qdisc->limit)>>2;
			 //exit slowstart phase
       if(slowstart[i] &&  backlog >= target)
         slowstart[i]=false;
			 //slowstart off =>inc/dec using the difference
       if(!slowstart[i])
         incr[i] += target - backlog;
       else 	//slowstart on =>add two segments
         incr[i] += 2 * MSS[i];
			 //update local window after M inc/dec(s)
       if (count == M)
       {
         localwnd[i]+=incr[i]/M;        
         localwnd[i]=MIN(localwnd[i], 65535 * conncount[i]);
         localwnd[i]=MAX(localwnd[i], TCP_MIN_MSS * conncount[i]);
         wnd[i]=localwnd[i]/conncount[i];
         incr[i]=0;
        }
        i++;
      }
      dev = next_net_device(dev);
    }
		//reset counter
    if(count == M)
      count=0;
    else
      count++;
		//if there are active connections rearm timer
    if(timerrun == true)
    {
      reschedule_timer(timer);
      return HRTIMER_RESTART;
    }
    else
      return HRTIMER_NORESTART;
}
\end{lstlisting}
\normalsize

\subsection{Packet arrival} 
As shown in listing~\ref{code:intercept}, for each new flow (again based on intercepted SYN-ACK packets), the current local window of the incoming and outgoing port is divided equally among all new active flows. However, for each torn down flow (based on intercepted FIN packets), the current window of the incoming port is redistributed equally among all currently active flows. Finally, If the ACK bit is set, the receive window field \Rwnd of this Packet is updated with the calculated per-port fair-share local window if it is smaller than the receive window value in TCP header.

\small
\begin{lstlisting}[caption=RNWDQ Packet Intercepter and Modifier, label=code:intercept]
//Packet interceptor and modifier
void rwndq_packet_arrival(sk_buff *skb, net_device *in, net_device *out)
{
	//in => index i and out => index j	
	//New connection setup
	if(tcp_header->syn && tcp_header->ack)
	{        
    conncount[i]+=1;//Increment connections of in
    conncount[j]+=1;//Increment connections of out
		if(conncount[i] >= 2)
      wnd[i] = wnd[i] * (conncount[i]-1) / conncount[i];
		if(conncount[j] >= 2)
      wnd[j] = wnd[j] * (conncount[j]-1) / conncount[j];
	}
	//Existing connection tear-down
	if(tcp_header->fin || tcp_header->rst)
  {
	  conncount[i]-=1;
    if(conncount[i] >= 1)      
      wnd[i]= wnd[i] * (conncount[i]+1) / conncount[i];
	}
	//Check for possible window modification
	if(tcp_header->ack) 
		if(wnd[i] < tcp_header->window)
		{
		 __be16 oldwnd = tcp_header->window;
		 tcp_header->window = wnd[i];
		 csum_replace2(tcp_header->check,oldwnd,wnd[i]);
		}           									
}
\end{lstlisting}
\normalsize

RWNDQ can maintain a very small buffer occupancies which allows the switch's small buffer to absorb transient traffic bursts while keeping the line busy. Therefore it achieves a high throughput for elephants and very small queuing delay and low loss probability for mice. 

RWQND as discussed uses proportional increase, proportional decrease rather than AIMD. Yet RWNDQ is still stable due to the locality of window control mechanism with the managed queues (i.e., local control loop). As soon as the queue occupancy increase above or decrease below the target threshold, the local window is shrunk or expanded in proportion, which ensures an average persistent queue occupancy level equal to target. Furthermore the increase/decrease amount is equally divided among all ongoing flows which ensures short and long term fairness among competing flows unlike end-to-end systems based on TCP.
 
\section{RWNDQ data center wide deployment}
\label{sec:3}

\begin{figure}[ht]
	\centering	 
		\includegraphics[scale=0.95]{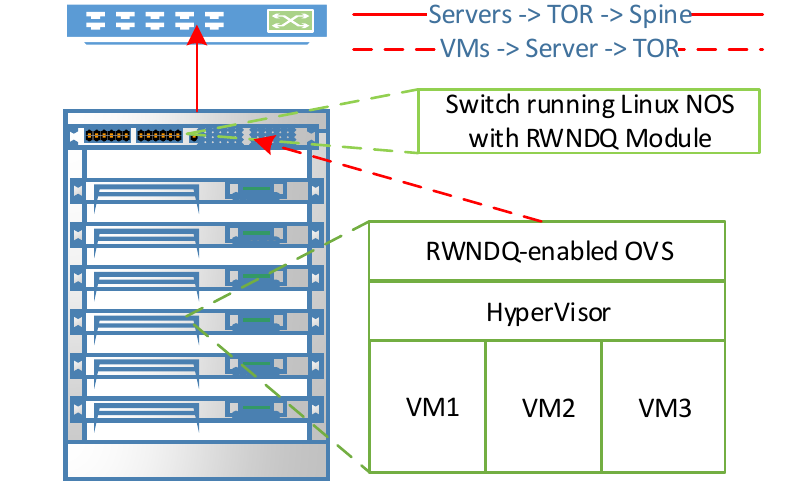}
	\caption{RWNDQ-enabled Data Center Architecture}
	\label{fig:torandservers}
\end{figure}

Because of the promising performance shown by RWNDQ in simulation studies, we have implemented the algorithm in a real testbed to ascertain and demonstrate its practical feasibility. Two options were implemented: 
\begin{itemize}
\item Linux Kernel Module: RWNDQ can be implemented using the Linux NetFilter framework \footnote{“NetFilter: Packet filtering framework for linux” http://www.netfilter.org/} as the NetFilter is a good candidate for intercepting incoming and outgoing packets. Also the NetFilter enables incrementing the protocol stack very easily via pre-registered hooks in the processing pipeline of the network stack; 
\item Open vSwitch: RWNDQ can also be implemented as an added feature to OvS by patching the OvS data-path kernel module to add RWNDQ's processing logic. 
\end{itemize}

Typically, a data center network consists of servers and switches interconnecting them. In oversubscribed data centers, the contention points in the network, are within the server on the outgoing interfaces between competing VMs and on the uplinks to upper layers (i.e the link from Top of Rack switch to the core switch) where multiple servers in a single rack share the uplink reach servers in other racks. In our RWNDQ system design, as shown in  Figure~\ref{fig:torandservers}, we propose to use hardware switches running Linux Network OS such as Open Networking Linux (ONL) \footnote{Open Network Linux http://opennetlinux.org} or PicOS \footnote{PicOS: http://www.pica8.com/white-box-switches/white-box-switch-os.php} on ToR and spine level hardware switches to interconnect servers within the racks. The switches with PicOS can support RWNDQ as a loadable kernel module or as a patched Open vSwitch. In addition, VMs within servers are interconnected via a software OvS which is the most popular choice for most cloud management frameworks like OpenStack.  RWNDQ as a kernel module and OvS-patch provide a potential for an easy deployment in production data centers at all different switching levels and possible congestion points in the network.
 
\subsection{RWNDQ as a loadable kernel module}
\label{subsec:rwndqnetfiler}

\begin{figure}[ht]
	\centering	 
		\includegraphics{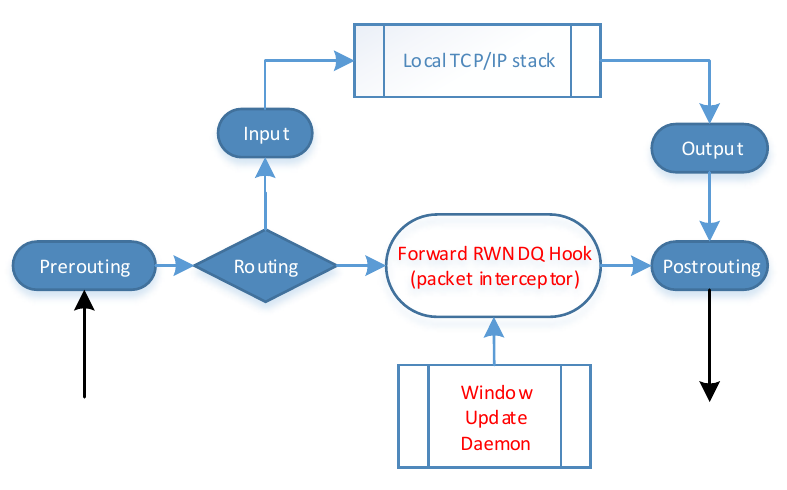}
	\caption{NetFilter-based RWNDQ packet processing pipeline}
	\label{fig:netfilterpipe}
\end{figure}

RWNDQ as a loadable kernel module is implemented using the NetFilter framework of the Linux operating system. In a non-virtualized environment, RWNDQ works as a hook that attaches to the data forwarding path in the Linux kernel just above the NIC driver and below the TCP/IP stack. This is a clean way of deploying RWNDQ which does not touch the TCP/IP implementation of the host operating system, making it easily deploy-able in production DCNs. In what follows, we introduce our Linux implementation which is used in this section's experiments. 

RWNDQ is implemented as a NetFilter hook as shown in Figure \ref{fig:netfilterpipe}. We insert the NetFilter hook at the forwarding stage of packet processing to intercept all forwarded TCP packets not destined to the host machine. Forwarding stage is executed right after the routing decision has been taken and immediately before the post-routing processing. As explained previously in the RWNDQ algorithm in section~\ref{sec:2}, TCP packet headers are examined and the processing is determined based on the SYN-ACK, FIN and ACK flag bits. In addition, since in DCNs transmission and propagation delays are in the microsecond time scale, Linux kernel timers based on the HZ tick rate traditionally used in the protocol stack and the OS, are not accurate enough to keep track of the queue occupancy as they are in the millisecond time scale; therefore, we invoke Linux high-resolution timers to deal with this\footnote{https://www.kernel.org/doc/Documentation/timers/hrtimers.txt}. The high resolution timer shown in the figure as ``Window udpate daemon'' is used to trigger switch's local per-port receiver window values updates based on the observed queue occupancy more accurately. The operations of the RWNDQ kernel module are described as follows:
\begin{itemize}
\item When a SYN-ACK packet is captured by the NetFilter hook, we increment the connection counter for both the ingress and egress Ethernet port and update their local window variables respectively\footnote{Note that, assuming current window value is in a stable state and optimal, by adding or deleting a TCP connection, the window value needs to be redistributed equally and fairly over the new number of active TCP flows}. Note that, SYN-ACK packets is sent only when TCP connection is established by the destination host of the connection and by default all TCP connections are full-duplex.
\item When a FIN packet is captured by the NetFilter hook, we decrement the connection counter for the ingress port only and update its local window variable. Note that, FIN packets are sent only when one side of the TCP connection finishes the transport of its application data and the other side of the TCP connection can still send data while the host which sent the FIN operates in half-closed state until it receives a FIN from its partner.
\item When an outgoing ACK packet is captured by the NetFilter hook, its receive window in TCP header will be checked against the local window, then the receiver window value is updated only if the local receive window value, pre-computed by the RWNDQ local window update mechanism, is smaller. 
\item If the window is updated, the checksum of the packet is  recomputed which is done using a kernel built-in function \textit{"csum\_replace2"}, which implements the update efficiently.
\item The high resolution timer is responsible for triggering the local window update function on  regular intervals, it is triggered on intervals smaller than the measured RTT in the network, in our setup RTTs are observed to be within a few hundred microseconds.
\item For each timer expiry, we calculate the increment value using the method described in RWNDQ algorithm in Section~\ref{sec:2}.
\item When the timer expires $M$ times consecutively, the local window value is updated using the accumulated increment values over the last $M$ intervals, the increment variable is reset and a new window update cycle starts from this point.
\end{itemize}

\subsection{Testbed setup}
\begin{figure}[ht]
	\centering	 
		\includegraphics[scale=0.7, height=3cm]{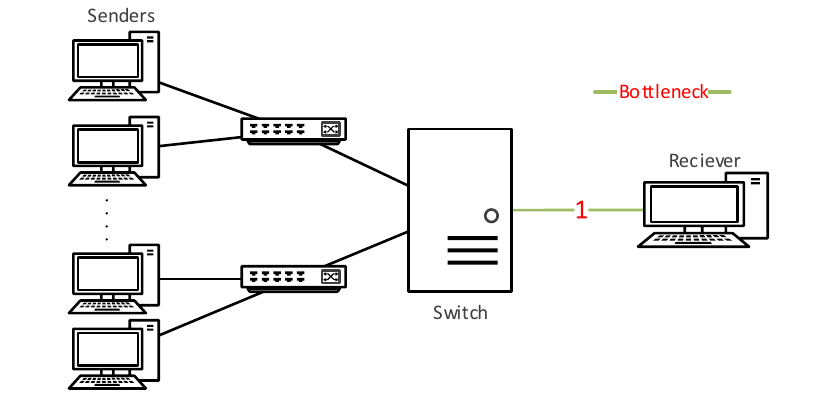}
	\caption{A dumbell-like setup to test RWNDQ Kernel Module}
	\label{fig:rwndqtestbed}
\end{figure}

To experiment with the Linux RWNDQ kernel module, we set up a single bottleneck testbed as shown in Figure~\ref{fig:rwndqtestbed}. The testbed consists of 6 Lenovo and 7 Dell desktops configured with core2duo processors and 4G of ram. One of the Dell desktops acts as the switch and it is equipped with 3 1Gbps Ethernet cards. %
In this setup, 5 machines and the switch are running Ubuntu 14.04 Desktop Edition with Linux kernel V3.13.0.34,  while the other 6 machines and the receiver (master) are running CentOS 6.6 with Linux kernel V3.10.63. All machines are running an Apache web server hosting the default small \textit{"index.html"} webpage of size 11.5KB. We rely on two well-known measurement applications for our experiments, iperf \cite{iperf} for generating elephants and Apache benchmark \cite{apacheb} for generating synchronized mice. The base RTT in our testbed is around $\approx$200$\mu$s. We allocate a static buffer size of 85.3KB to all ports in the network using Linux Traffic Control (Linux TC). In all experiments, we set up the queuing discipline \textit{Qdisc} of each Ethernet port in the network to \textit{bfifo} queue with limit of 85.3KB bytes. This value matches the buffer size for each port in a switch like pronto 3295, that has 4MB of shared buffer memory used by 48 ports, leaving a buffer of 85.3KB for each port on the switch. We evaluated the experiments with both cubic TCP and new-reno TCP, the only available congestion control mechanisms in both Linux kernel versions (3.10 and 3.13).              

\subsection{Experimental Results}

Eleven iperf traffic flows (elephants) are started at the same time from each of the senders towards the receiver which is connected to one of the switch ports.%
The flows send continuously for 50 secs and throughput samples are collected over 0.5 secs intervals. At the $20^{th}$ second, each sender starts Apache Benchmark to request \textit{"index.html"} 1000 times (mice) and report statistics on the completion times.

\begin{figure}[ht]
        \centering        
        \begin{subfigure}[ht]{0.49\columnwidth}
            \includegraphics[scale=0.99, width=\textwidth]{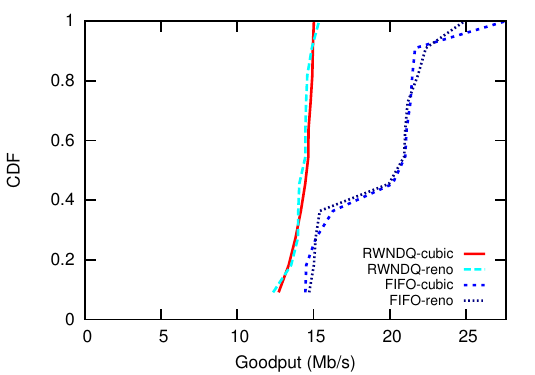}
              \caption{CDF of average throughput}
              \label{fig:module-50-1-cdf}
        \end{subfigure}
        \begin{subfigure}[ht]{0.49\columnwidth}
            \includegraphics[scale=0.99, width=\textwidth]{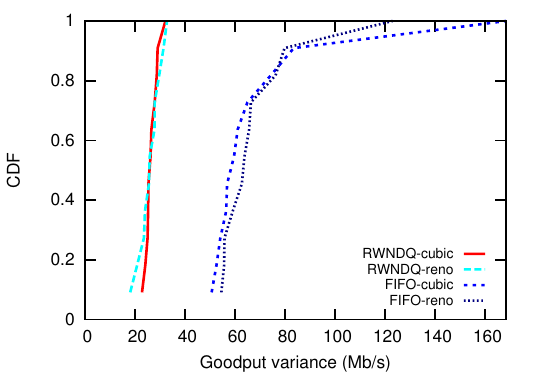}
                \caption{CDF of throughput variance}
                \label{fig:module-50-1-variance-cdf}
        \end{subfigure}
         \begin{subfigure}[ht]{0.49\columnwidth}
            \includegraphics[scale=0.99, width=\textwidth]{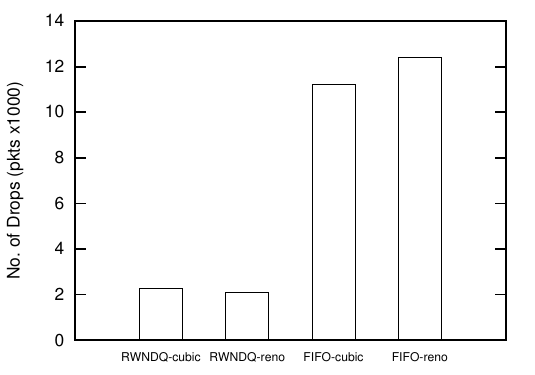}
              \caption{Total drops of the bottleneck link}
              \label{fig:module-50-1-drops}
        \end{subfigure}
				\caption{Elephant TCP flows performance: RWNDQ vs FIFO in a the Linux testbed with one bottleneck-link with new-reno and cubic TCP sources}
				\label{fig:module-bloat-11senders-good}
\end{figure}

Figure~\ref{fig:module-50-1-cdf} and ~\ref{fig:module-50-1-variance-cdf} show that RWNDQ helps both TCP variants to achieve a very close fair-share throughput to TCP-FIFO yet reduces the variations of the reclaimed throughput during elephant sessions, even with sudden surges of mice traffic. Figure~\ref{fig:module-50-1-drops} shows RWNDQ's ability to significantly reduce packet drops at the bottleneck link by $\approx80-85\%$. 

\begin{figure}[ht]
\captionsetup[subfigure]{justification=centering}
\centering
				\begin{subfigure}[ht]{0.49\columnwidth}
      \includegraphics[scale=1.1, width=\textwidth]{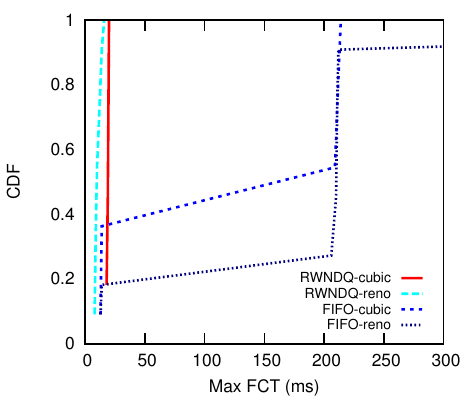}
                \caption{Max FCT}
                \label{fig:module-50-1-100-cdf}
        \end{subfigure}
        \begin{subfigure}[ht]{0.49\columnwidth}
      \includegraphics[scale=1.1, width=\textwidth]{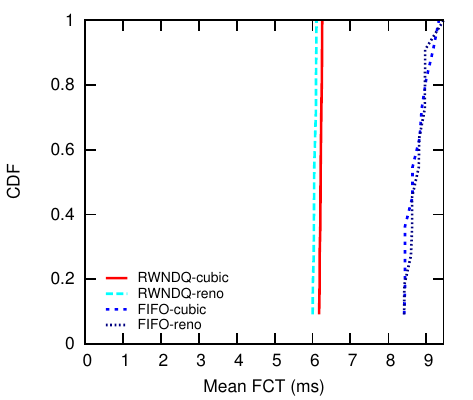}
                \caption{Average FCT}
                \label{fig:module-50-1-mean-cdf}
        \end{subfigure}				
				\caption{Mice TCP flows performance: RWNDQ vs FIFO in a the Linux testbed with one bottleneck-link with new-reno and cubic}
				\label{fig:module-bloat-11senders-time}
\end{figure}

The reduction of packet drops benefits mice flows by avoiding unnecessary timeouts. As Figure~\ref{fig:module-50-1-100-cdf} shows, mice tail flow completion time (FCT) is less than 200ms, which is the default $RTO_{Min}$ in Linux. Finally, according to Figure~\ref{fig:module-50-1-mean-cdf}, RWNDQ allows competing mice to finish quickly and at approximately the same time.

\section{RWNDQ-enabled Open vSwitch}
\label{sec:4}

We further investigated the implementation of RWNDQ in OvS as this latter already implements SDN based flow tracking. We patched the Kernel data-path modules of OvS with the same functions described earlier in the RWNDQ Linux-kernel module. In this case however, we did not use the NetFilter hook, we instead added RWNDQ functions in the processing pipeline of the packets that pass through the kernel datapath module of OvS. In a virtualized environment, RWNDQ-enabled OvS can process the traffic for inter-VM, Intra-Host and Inter-Host communications. This is an efficient way of deploying RWNDQ  on the host operating system of the switch by only applying a patch and recompiling OvS module, making it easily deploy-able in today's production DCs.

\begin{figure}[ht]
\captionsetup[subfigure]{justification=centering}
\centering
			\begin{subfigure}[ht]{0.8\columnwidth}
            \includegraphics[width=\columnwidth]{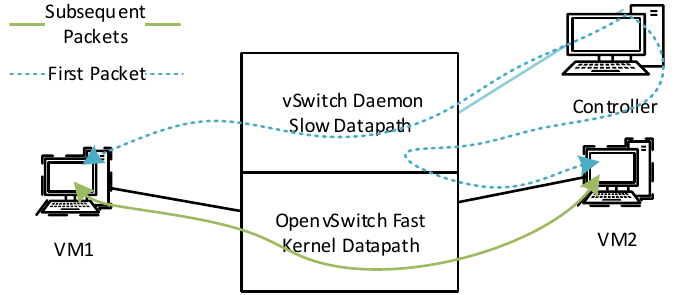}             
						\caption{OpenvSwitch slow and fast datapath}
             \label{fig:ovsarch}
       \end{subfigure}
				\\
			\begin{subfigure}[ht]{0.8\columnwidth}
       \includegraphics[width=\columnwidth]{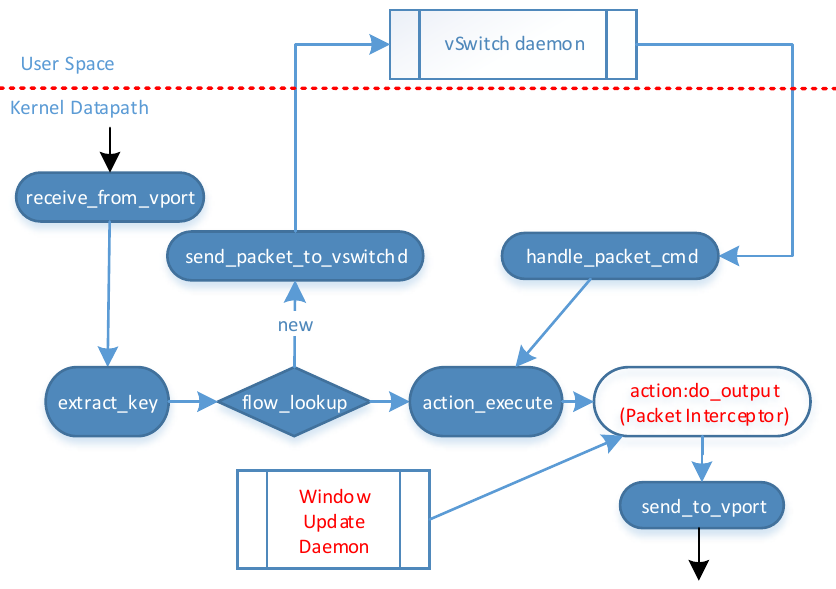}
                 \caption{OpenvSwitch-based RWNDQ packet processing pipeline}
                \label{fig:ovspipe}
        \end{subfigure}				
				\caption{RWNDQ OpenvSwitch-based implementation}
				\label{fig:openvswitch}
\end{figure}

As show in Figure~\ref{fig:openvswitch}, OvS is mainly composed of two parts, the data path kernel module and the user-space vSwitch daemon that communicates with the controller using OpenFlow protocol over encrypted SSH connections. OvS is flow-aware by design and all flow decision entries in the forwarding table are inserted by a local or a remote controller. Whenever, a packet arrives at any port of the switch, its flow key is hashed and examined against current active flows in the table. If the entry for that flow could not be found, the packet is immediately forwarded to the controller for establishing the identity of this flow and setting up the forwarding entries in all involved switches of the network. Primarily, any packet is processed by the kernel fast data-path only if its flow entry is active in the forwarding table, in such case, the packet is forwarded immediately without experiencing any further delays. RWNDQ's packet interceptor and its packet handling logic described in Section~\ref{subsec:rwndqnetfiler} are inserted in processing of $do\_output$ action function as shown in Figure~\ref{fig:ovspipe}. TCP packets being forwarded are intercepted and their window is updated if necessary. Local per-port window values are updated on a regular basis by the window update daemon. 

\subsection{Testbed Setup}
\begin{figure}[ht]
	\centering	 
		\includegraphics[scale=0.7, height=3cm]{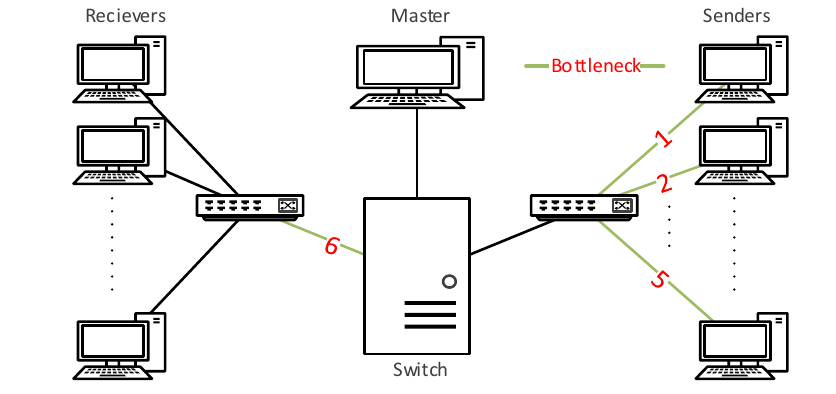}
	\caption{Small testbed for RWNDQ-enabled OvS}
	\label{fig:cloudtestbed}
\end{figure}

For experimenting with our patched OvS, we set up a testbed as shown in Figure~\ref{fig:cloudtestbed}, it is similar to the testbed in Section~\ref{sec:3}, however, in this case, all machines' internal ports are connected to the patched OvS and the CentOS and Ubuntu hosts are connected to different 1 Gb/s D-Link dumb-switch. Here also, different scenarios are set up to reproduce both incast and buffer-bloating situations, however in this case multiple-bottleneck links in the network exist as shown in Figure~\ref{fig:cloudtestbed}. The bottlenecks at the senders are created by creating multiple ports on the OvS and binding an iperf or an Apache process to each one of them.  

\subsection{Experimental Results}

The goals of the experiments are to: 
\begin{inparaenum}[\itshape i) \upshape]
\item show that with the support of RWNDQ, TCP can support many more connections and maintain high link utilization; 
\item show that with the support of RWNDQ, TCP can overcome incast congestion situations in the network;
\item measure RWNDQ's impact on the FCT of mice flows; and,
\item explore RWNDQ's performance in buffer-bloating situations where mice compete with elephants.
\end{inparaenum}

\subsubsection{Incast Scenarios}
\begin{figure}[ht]
\captionsetup[subfigure]{justification=centering}
\centering
        \begin{subfigure}[ht]{0.49\columnwidth}
            \includegraphics[width=\textwidth]{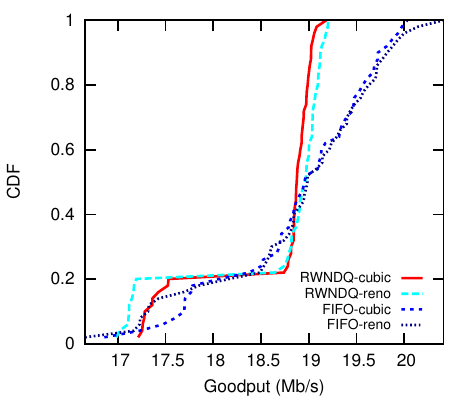}
              \caption{CDF of average throughput}
              \label{fig:incast-50-10-cdf}
        \end{subfigure}
        \begin{subfigure}[ht]{0.49\columnwidth}
            \includegraphics[width=\textwidth]{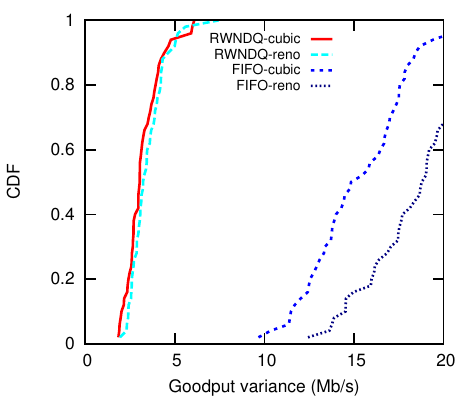}
                \caption{CDF of throughput variance}
                \label{fig:incast-50-10-varcdf}
        \end{subfigure}
				\caption{Comparison of TCP's performance with RWNDQ vs. FIFO with 50 elephants}\label{fig:incast-50senders}
				\label{fig:incast-10senders}

\end{figure}

We produce an incast-like scenario with synchronized senders all converging to the same output port resulting in excessive pressure on the output buffer in links 1-5 as well as link 6. First, we generated 10 iperf clients at same time from each of the 5 senders  destined to a separate iperf server listening on a separate port on the receivers. This results in 50 senders continuously sending for 50 secs and iperf is set to generate the throughput samples over 0.5 sec intervals. In the following we show the CDF of the average achieved throughput, the variance of the throughput samples and the total packet drops experienced at the bottleneck links during the experiment. Figure~\ref{fig:incast-50senders} shows that in a medium load situation, RWNDQ helps TCP in achieving a balanced distribution of bandwidth among competing senders and reduces the variations of their reclaimed bandwidth during the lifetime of a TCP connection. Table~\ref{tab:10-drops} clearly shows how RWNDQ switch queue management is able to reduce the number of packet drops at bottleneck links by $\approx92-99\%$ (nearly two orders of magnitude), reducing considerably unnecessary timeouts for TCP connections and allowing the flows to finish at approximately the same time. This due to RWNDQ dividing the effective window (bandwidth-delay product + target buffer length) equally among competing flows, allowing them to achieve nearly the same throughput and hence very similar flow completion times.

\begin{table*}[!t]
\centering
\small
\caption{Number of packet drops experienced at each of the 6 bottleneck links labeled 1 to 6 in Figure~\ref{fig:cloudtestbed}}
\begin{subtable}{.48\linewidth}
\centering
\small
\caption{50 elephants scenario}
\begin{tabular}{@{}c|c|c|c|c|@{}}
\cline{2-5}
\multicolumn{1}{c|}{}		& \multicolumn{2}{c|}{Reno} & \multicolumn{2}{c|}{Cubic} \\ \cline{2-5} 
 \multicolumn{1}{c|}{}  & RWNDQ        & FIFO       & RWNDQ        & FIFO        \\ \hline
\multicolumn{1}{|c|}{1} & 10           & 4992       & 33            & 4605        \\ \hline
\multicolumn{1}{|c|}{2} & 5            & 4913       & 21           & 4548        \\ \hline
\multicolumn{1}{|c|}{3} & 10           & 4676       & 19            & 4319        \\ \hline
\multicolumn{1}{|c|}{4} & 18           & 4860       & 29           & 4530        \\ \hline
\multicolumn{1}{|c|}{5} & 12           & 4857       & 44            & 4520        \\ \hline
\multicolumn{1}{|c|}{6} & 531          & 331        & 320            & 357         \\ \hline
\end{tabular}
\label{tab:10-drops}
\end{subtable}
\begin{subtable}{.48\linewidth}
\centering
\small
\caption{200 elephants scenario}
\begin{tabular}{@{}c|c|c|c|c|@{}}
\cline{2-5} 
 \multicolumn{1}{c|}{}	& \multicolumn{2}{c|}{Reno} & \multicolumn{2}{c|}{Cubic} \\ \cline{2-5} 
 \multicolumn{1}{c|}{}  & RWNDQ        & FIFO       & RWNDQ        & FIFO        \\ \hline
\multicolumn{1}{|c|}{1} & 1750           & 30934       & 2184            & 30422        \\ \hline
\multicolumn{1}{|c|}{2} & 1671           & 30851       & 2361           & 30767        \\ \hline
\multicolumn{1}{|c|}{3} & 2544           & 27486       & 2418            & 28276        \\ \hline
\multicolumn{1}{|c|}{4} & 1632           & 30620       & 2152           & 30210        \\ \hline
\multicolumn{1}{|c|}{5} & 1547           & 30860       & 2249            & 30540        \\ \hline
\multicolumn{1}{|c|}{6} & 3394          & 12901        & 3516            & 23432         \\ \hline
\end{tabular}
\label{tab:40-drops}
\end{subtable}
\end{table*}
\normalsize

\begin{figure}[ht]
\captionsetup[subfigure]{justification=centering}
\centering
        \begin{subfigure}[ht]{0.49\columnwidth}
            \includegraphics[scale=0.99, width=\textwidth]{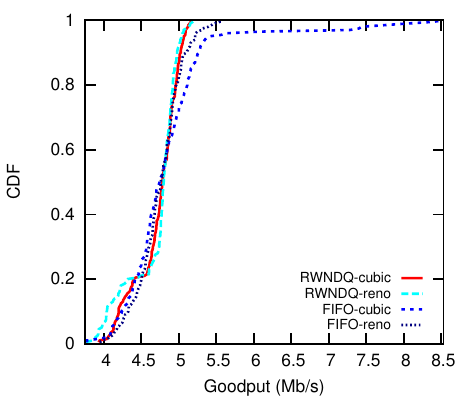}
              \caption{CDF of average throughput}
              \label{fig:incast-50-40-cdf}
        \end{subfigure}
        \begin{subfigure}[ht]{0.49\columnwidth}
            \includegraphics[scale=0.99, width=\textwidth]{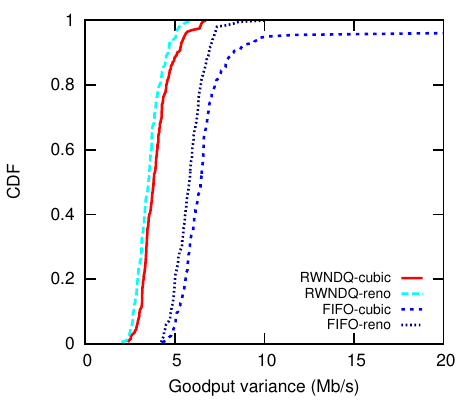}
                \caption{CDF of throughput variance}
                \label{fig:incast-50-40-varcdf}
        \end{subfigure}
				\caption{Comparison of TCP's performance with RWNDQ vs. FIFO with 200 elephants}
				\label{fig:incast-40senders}
\end{figure}

We repeat the same experiment but this time we increase the number of iperf (elephant) senders per host to 40 (resulting in a total 200 elephants). Again, Figure~\ref{fig:incast-40senders} supports our claims, and shows that even in a high load, RWNDQ still helps TCP (new-reno and cubic) achieve a balanced distribution of bandwidth and maintains a very low variation of throughput for TCP connections involved in the incast. We observe that the variation for some TCP flows without RWNDQ reaches $\approx$400Mbps, the reason being, for some time intervals, a few flows grab most of the bandwidth while the others achieve nearly zero throughput. Table~\ref{tab:40-drops} shows that RWNDQ is still able to keep a very low packet drop rate by one order of magnitude compared to TCP without the assistance of RWNDQ mechanism at the switch.

\subsubsection{Buffer-Bloating Scenarios}

\begin{figure}[ht]
\captionsetup[subfigure]{justification=centering}
\centering
			\begin{subfigure}[ht]{0.49\columnwidth}
            \includegraphics[width=\textwidth]{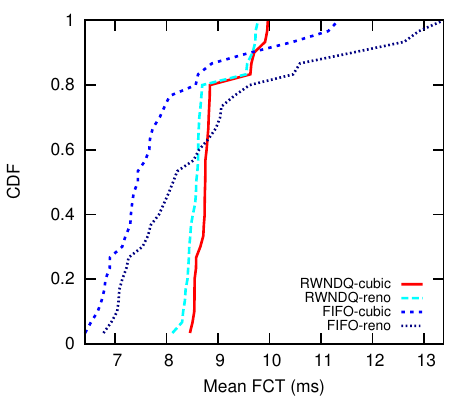}             
						\caption{Average mice FCT}
                \label{fig:bloat-50-10-mean-cdf}
       \end{subfigure}
			\begin{subfigure}[ht]{0.49\columnwidth}
            \includegraphics[width=\textwidth]{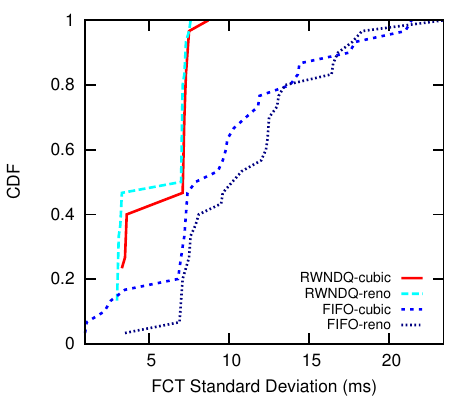}             
						\caption{SD of mice FCT}
              \label{fig:bloat-50-10-sd-cdf}
       \end{subfigure}
				\\
			\begin{subfigure}[ht]{0.49\columnwidth}
       \includegraphics[width=\textwidth]{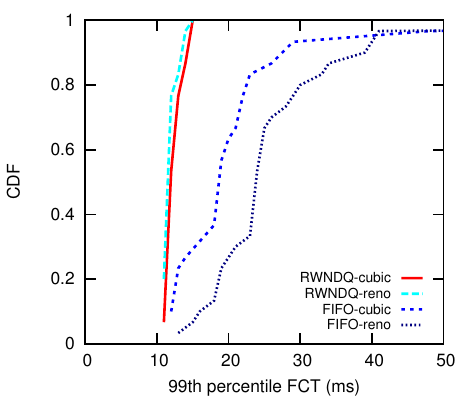}
                 \caption{$99^{th}$ percentile of mice FCT}
                \label{fig:bloat-50-10-99-cdf}
        \end{subfigure}				
			  \begin{subfigure}[ht]{0.49\columnwidth}
       \includegraphics[width=\textwidth]{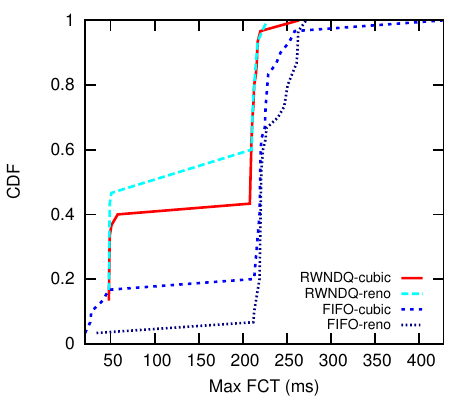}
                 \caption{Max mice FCT}
                \label{fig:bloat-50-10-100-cdf}
        \end{subfigure}
				\caption{Comparison of mice FCT for RWNDQ vs. FIFO where 50 elephants compete with 30 mice}
				\label{fig:bloat-10senders}
\end{figure}

We reproduce a buffer-bloating scenario in which mice traffic compete with elephant flows to see if RWNDQ can reconcile the two classes. Similar to the previous experiment, we first generate 10 synchronized iperf elephant connections continuously sending for 50 secs from each sender resulting in 50 elephants at link 6. We use Apache benchmark to request \textit{"index.html"} webpage (representing mice flows) from each of the web servers ($6\times5=30$ in total) running on the same machines where elephants are sending. Note that, we run Apache benchmark, at the $20^{th}$sec, requesting the webpage 1000 times then it reports different statistics over the 1000 requests. The performance of elephants was close to what has been presented in the incast scenario experiments. Now, Figure~\ref{fig:bloat-10senders} shows that, in medium load, RWNDQ achieves a good balance in meeting the conflicting requirements of elephants and mice. The competing mice flows benefit under RWNDQ by achieving a nearly equal FCT on average with very small standard deviation compared to TCP with FIFO as shown in Figure~\ref{fig:bloat-50-10-mean-cdf} and~\ref{fig:bloat-50-10-sd-cdf}. In addition, as RWNDQ efficiently regulates the flows and keeps the drop rate near to zero, in Figure~\ref{fig:bloat-50-10-99-cdf}, the $99^{th}$ percentile for RWNDQ never crosses the 200ms threshold which is the default $RTO_{min}$ of Linux, as opposed to TCP with FIFO which can be attributed to the timeouts caused by high drop rates. In Figure~\ref{fig:bloat-50-10-99-cdf}, the maximum FCT  $\approx40-60\%$ of the flows are below the 200ms with RWNDQ compared to only $\approx1-18\%$ for TCP with FIFO.

\begin{figure}[ht]
\captionsetup[subfigure]{justification=centering}
\centering
				\begin{subfigure}[ht]{0.49\columnwidth}
        \includegraphics[scale=1.1, width=\textwidth]{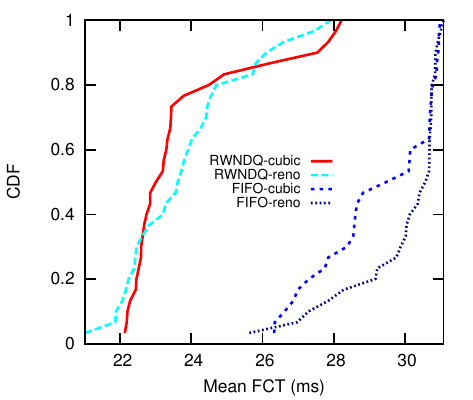}
                \caption{Average mice FCT}
                \label{fig:bloat-50-40-mean-cdf}
        \end{subfigure}
				\begin{subfigure}[ht]{0.49\columnwidth}
        \includegraphics[scale=1.1, width=\textwidth]{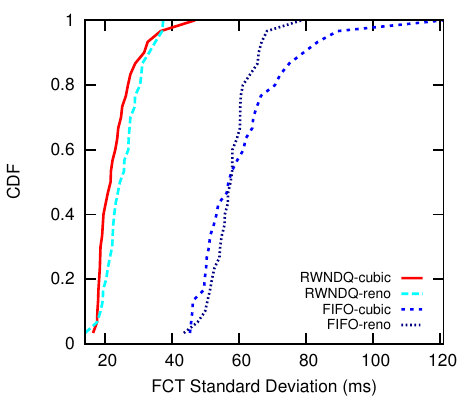}
                \caption{SD deviation of mice FCT}
                \label{fig:bloat-50-40-sd-cdf}
        \end{subfigure}
				\\
					\begin{subfigure}[ht]{0.49\columnwidth}
           \includegraphics[scale=1.1, width=\textwidth]{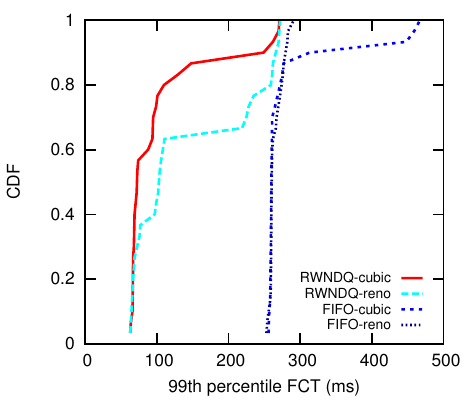}
               \caption{$99^{th}$ percentile mice FCT}
                \label{fig:bloat-50-40-99-cdf}
        \end{subfigure} 
				\begin{subfigure}[ht]{0.49\columnwidth}
           \includegraphics[scale=1.1, width=\textwidth]{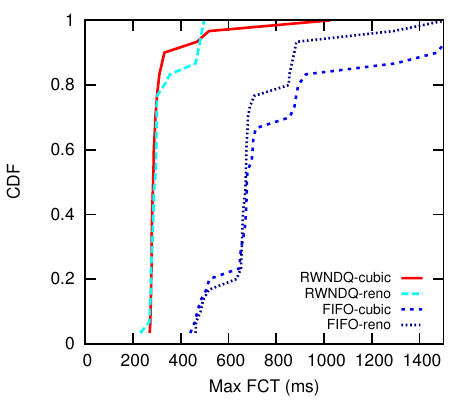}
               \caption{Max mice FCT}
                \label{fig:bloat-50-40-100-cdf}
        \end{subfigure} 
				\caption{Comparison of mice FCT with RWNDQ vs. FIFO where 200 elephants compete with 30 mice}
				\label{fig:bloat-40senders}
\end{figure}

Again, we repeat the high load experiment with 200 elephants and introduce the 30 competing mice. As shown in Figure~\ref{fig:bloat-40senders}, RWNDQ is able to satisfy the requirements of latency-sensitive mice even-though they are outnumbered by elephants. Figure~\ref{fig:bloat-50-40-mean-cdf} and~\ref{fig:bloat-50-40-sd-cdf} show that mice flows are not blocked by the bandwidth-hogging elephants. The mean FCT under RWNDQ are small and the CDF curve is smooth, according to the standard deviation, in contrast to what is achieved with FIFO. In addition, Figure~\ref{fig:bloat-50-40-99-cdf} and~\ref{fig:bloat-50-40-100-cdf} show that, the tail and the $99^{th}$ percentile of the FCT of TCP with FIFO is experiencing timeouts as indicated by FCT values of over 250ms. Meanwhile RWNDQ avoids timeouts by managing the queue efficiently and hence it greatly reduces the FCT of mice on the tail $99^{th}$ percentile on average by $\approx60\%$. 

\subsubsection{RWNDQ system overhead}
To quantify system overhead introduced by the RWNDQ packet intercepting and modifying module, we measured CPU usage on the server operating as the switch between the other servers which is equipped with Intel Core2 Duo CPU running at 2.13GHz and 4 GBytes of Ram. We rerun the high load experiment with 200 elephants and the 30 competing mice. We achieved a high link utilization of $\approx$~900-935 Mbps goodput while the extra CPU usage introduced by RWNDQ is $\approx~1\%$ compared with the case where the RWNDQ module is not enabled.

\subsubsection{Summary of the experimental Results}

In summary the experimental results reinforce the results obtained in the simulation study conducted in \citep{Ahmed-CLOUDNET-2015}. In particular, they show that:
\begin{itemize}
\item RWNDQ helps in reducing mice traffic latency and maintains sufficient throughput for elephants. In the experiments both elephants and mice are able to achieve their requirements as stated in the Introduction.
\item RWNDQ can easily handle congestion, in low to high load incast or buffer-bloating scenarios, while nearly saturating the link at rate of $\approx$~900-935 Mbps, which matches our findings from our NS-2 simulations.
\item RWNDQ achieved all this without any modification to the TCP congestion control mechanism at the source nor to the receiver and seems to scale well in our testbed.
\end{itemize}

\section{Related Work}
\label{sec:5}

Many recent work have been devoted to addressing shortcomings of TCP in data centers. The proposed mechanisms can be classified into:
\begin{enumerate}
\item window-based schemes (e.g., \cite{Ahmed-CLOUDNET-2015,Ahmed-LCN-2017,Alizadeh2010,Wu2013, Ahmed-GLOBECOM-2018, Ahmed-GLOBECOM-2015,Ahmed-INFOCOM-2019}). For instance, DCTCP \cite{Alizadeh2010} uses a series of ECN-feedback information to adjust the TCP congestion window to stabilize the queue length in the switch at a predefined small threshold. ICTCP \cite{Wu2013} proposed modifies the TCP receiver in order to handle incast traffic by proactively adjusting the TCP receiver window to avoid congestion at the receiver. RWNDQ \cite{Ahmed-CLOUDNET-2015} proposed to not modify the TCP end-points and instead modifies the switch instead to set the target rate in the receive window field to ensure fair allocation of network capacity among the competing flows. Several follow-up work leveraged SDN towards solving the problem~\cite{Ahmed-ICC-2017,Ahmed-ICC-2016-1,Ahmed-ANNALS-2017,Ahmed-LCN-2017} or applied same ideas in the context content centric networks~{Ahmed-ICC-2016-2,Ahmed-LCN-2016}

\item fast loss recovery schemes (e.g., \cite{Vasudevan2009, Cheng2014, Ahmed-INFOCOM-2018, Ahmed-ICDCS-2019-1, Ahmed-ICPP-2020,Ahmed-TON-2021}). These schemes try to shorten the reaction time to enhance the recovery mechanisms of TCP in response to congestion events. For instance, \cite{Vasudevan2009,Ahmed-INFOCOM-2018,Ahmed-TON-2021} adjusts the minimum RTO $RTO_{min}$ of TCP flows in data centers which result in faster recovery times from congestion losses. \cite{Cheng2014} propose a mechanism to truncate the payload of  packets causing congestion in the network and forwarding only the header to the receiver. Other approaches consider the co-flow abstraction to collectively optimize the performance of flows who share the same goal or task~\cite{Ahmed-ICC-2019,Ahmed-ICDCS-2019-2}.
\end{enumerate}

In this work, because RWNDQ~\cite{Ahmed-CLOUDNET-2015,Ahmed-IPCCC-2015} showed significant improvements over the other methods in different simulation scenarios, we are set to make a practical case for RWNDQ by designing, implementing, and evaluating a working prototype of it. To achieve our goal, the switch is modified track the number of ongoing flows and accordingly feedback the fair share allocation of the bandwidth of each back their sources. This scheme induces low overhead on the switch (i.e., minimal flow state information) and requires a simple update to switch software to allow it to rewrite the receiver window field in TCP header. Additionally, it requires no modifications to the networking stack (i.e., TCP) of the end-hosts.

\section{Conclusion and future work}
\label{sec:6}

In this paper, we set to demonstrate the implementation for RWNDQ mechanism and its immediate operability in data center networks. RWNDQ is designed to reconcile the non-compatible requirements of elephant flows and mice flows that account for the majority of datacenter traffic. RWNDQ is designed to maintain a small persistent queue size to leave room in the buffer to absorb sudden transient bursts of incast traffic. Hence, it can decrease the average flow completion time of mice flows, yet maintain a high throughput for elephants. RWNDQ is a switch-assisted congestion-control system that builds on top of the existing flow-control of TCP to feedback queue occupancy levels to TCP senders. RWNDQ is designed to avoid any modification to the VM TCP protocol as a result it can be adopted easily for public DC networks where different TCP variations co-exist. To prove the realistic feasibility of our approach, we implemented RWNDQ as a standalone Linux kernel module easily deployable on hardware switches running a Linux network OS or as an added feature to the well known open vSwitch kernel data-path module for deployment in current virtualized DC networks. The results of our experiments strongly suggest that a switch-based approach like RWNDQ is a good approach to handle incast and buffer-bloating situations simultaneously. As part of our future work, we aim to implement a NetFPGA based prototype to have a real hardware switch~\cite{Ahmed-ITCE-2019}.

\balance
\bibliography{paper9,online}
\bibliographystyle{ieeetr}
\balance
\end{document}